\documentclass[smallextended]{svjour3}       % onecolumn (second format)
\smartqed  % flush right qed marks, e.g. at end of proof
\usepackage{graphicx}
\usepackage{setspace}
\usepackage{amsmath,amssymb}
\usepackage{mathrsfs}
\usepackage{bm}
\usepackage{color}

\begin{document}

\title{Open quantum dynamics theory on the basis of periodical system-bath model for discrete Wigner function\thanks{Y.~T.~is supported by JSPS KAKENHI Grant Number B 21H01884}
}

\titlerunning{Open quantum dynamics theory  for discrete Wigner function }        % if too long for running head

\author{Yuki Iwamoto         \and
        Yoshitaka Tanimura %etc.
}

\institute{Yuki Iwamoto \at
              Department of Chemistry, Graduate School of Science, Kyoto University, Kyoto 606-8502, Japan \\
              \email{iwamoto.yuki.w57@kyoto-u.jp}           %  \\
           \and
           Yoshitaka Tanimura \at
              Department of Chemistry, Graduate School of Science, Kyoto University, Kyoto 606-8502, Japan \\
              \email{tanimura.yoshitaka.5w@kyoto-u.jp}  
}

\date{Received: date / Accepted: date}

\maketitle

\begin{abstract}
Discretizing a distribution function in a phase space for an efficient quantum dynamics simulation is a non-trivial challenge, in particular for a case that a system is further coupled to environmental degrees of freedom. Such open quantum dynamics is described by a reduced equation of motion (REOM) most notably by a quantum Fokker-Planck equation (QFPE) for a Wigner distribution function (WDF). To develop a discretization scheme that is stable for numerical simulations from the REOM approach, we employ a two-dimensional (2D) periodically invariant system-bath (PISB) model with two heat baths. This model is an ideal platform not only for a periodic system but also for a non-periodic system confined by a potential. We then derive the numerically ''exact'' hierarchical equations of motion (HEOM) for a discrete WDF in terms of periodically invariant operators in both coordinate and momentum spaces.  The obtained equations can treat non-Markovian heat-bath  in a non-perturbative manner at finite temperatures regardless of the mesh size. As demonstrations, we numerically integrate the discrete QFPE  for a 2D free rotor and harmonic potential systems in a high-temperature Markovian case using a coarse mesh with initial conditions that involve singularity.

\end{abstract}

\section{Introduction}
\label{intro}
A central issue in the development of a computational simulation for a quantum system described in a phase space distribution is the instability of the numerical integration of a kinetic equation in time, which depends upon a discretization scheme of the coordinate and momentum\cite{Frensley,Jacoboni2004,Grossman,Kim2007,Weinbub}. In this paper, we introduce a new approach in order to construct a Wigner distribution function (WDF) for an open quantum dynamics system on the basis of a finite-dimensional quantum mechanics developed by Schwinger \cite{Schwinger}. Here, open quantum dynamics refers to the dynamics of a system coupled to baths consisting of surrounding atoms or molecules that is typically modeled by an infinite number of harmonic oscillators \cite{Caldeira-AP-1983,Weiss08,Breuer,CaldeiraPhysica83,WaxmanLeggett1985,Tanimura89A,TanimuraPRA91,TanimuraJCP92,Tanimura-2006,YTperspective,YTJCP2014,YTJCP2015}. After reducing the bath-degrees of freedom, the derived reduced equation of motion can describe the time irreversibility of the dynamics toward the thermal equilibrium state. The energy supplied by fluctuations and the energy lost through dissipation are balanced in the thermal equilibrium state, while the bath temperature does not change, because its heat capacity is infinite. 

In previous studies, the Boltzmann collision operator \cite{JensenPRL1991,JensenIEEE1991} and the Ornstein--Uhlenbeck operator \cite{CaldeiraPhysica83,WaxmanLeggett1985} have been used for a description of dissipative effects in the quantum Boltzmann equation and quantum Fokker-Planck equation (QFPE), respectively. The former one, however, is phenomenological \cite{Zhan2016}, whereas the latter one is valid only at high temperature \cite{TanimuraPRA91} that  leads to a breakdown of the positivity of population distributions at low temperature \cite{Tanimura-2006,YTperspective,YTJCP2014}. This is because a Markovian assumption cannot take into account the effects of quantum noise, which is non-Markovian at low temperature. Thus, numerically ``exact'' approach, for example quantum hierarchical Fokker-Planck equations (QHFPE) \cite{YTJCP2015} for a reduced WDF must be used as the rigorous quantum mechanical treatments. These equations are derived on the basis of the hierarchical equation of motion (HEOM) formalism \cite{Tanimura89A,Tanimura-2006,YTperspective}. By using the QHFPE, for example, self-excited currentoscillations of the resonant tunneling diode (RTD) in the negative differential resistance region described by a Caldeira-Leggett model was discovered in a numerically rigorous manner  \cite{SakuraiJPSJ13,SakuraiNJP14,GrossmanSakuraiJPSJ}.

For a case of isolated time-reversible processes, a finite-difference approximation  of momentum operators  allows us  to solve a kinetic equation using a uniformly spaced mesh in the coordinate space. Then the wave function is expressed in this discretized space. While the quantum dynamics of an isolated $N$-discretized coordinate system are described using the wave function as a $N$-dimensional vector, an open quantum system must be described using a $N \times N$ reduced density matrix, most notably for the quantum master equation (QME) approach or a $N \times M$ WDF for the $N$-discretized coordinate and the $M$-discretized momentum, most notably for the QFPE approach \cite{Frensley}.

Whether a system is isolated or is coupled to a heat bath, the WDF is numerically convenient and physically intuitive to describe the system dynamics. This is because the WDF is a real function in a classical phase like space and the described wavepacket in the momentum space is localized in a Gaussian like form following the Boltzmann statics, while the distribution in the coordinate space is spread. Various numerical schemes for the WDF, including the implementation of boundary conditions, for example inflow, outflow, or absorbing boundary conditions \cite{Ringhofer1989,Jiang2014,Mahmood2016}, and a Fourier based treatment of potential operators \cite{Frensley,Grossman}, have been developed. Varieties of application for quantum electronic devices \cite{Tsuchiya2009,Morandi2011,Barraud2011,Jonasson2015,Tilma2016,Ivanov2017,Kim2016,Kim2017}, most notably the RTD \cite{Frensley1986,Frensley1987,FerryPRB1989,Ferry2003,Buot1991,BuotPRL1991,Buot2000,Biegel1996,Yoder2010,Jiang2011,Dorda2015} that includes the results from the QHFPE approach \cite{SakuraiJPSJ13,SakuraiNJP14,GrossmanSakuraiJPSJ}, quantum ratchet \cite{Zueco,Coffey2009,KatoJPCB13}, chemical reaction \cite{TanimuraPRA91,TanimuraJCP92}, multi-state nonadiabatic electron transfer dynamics \cite{MS-QFP94,MS-QFPChernyak96,TanimuraMaruyama97,MaruyamaTanimura98,Ikeda2019Ohmic,Ikeda2017JCP}, photo-isomerization through a conical intersection \cite{Ikeda2018CI}, molecular motor \cite{Ikeda2019JCP}, linear and nonlinear spectroscopies \cite{SakuraiJPC11,Ikeda2015,Ito2016}, in which the quantum entanglement between the system and bath plays an essential role, have been investigated. 

The above mentioned approaches have utilized a discrete WDF. Because original equations defined in continuum phase space are known to be stable under a relevant physical condition, any instability arises from a result of the discretization scheme.  In principle, discussions for a stability of the scheme involve a numerical accuracy of the discretization scheme with respect to the coordinate and momentum. Generally, the stability becomes better for finer mesh. However, computational costs become 
expensive and numerical accuracy becomes worse if the mesh size is too small. In addition, when the mesh size is too large, the computed results diverge as a simulation time goes on. Thus, we have been choosing the mesh size to weigh the relative merits of numerical accuracy and costs.

In this paper, we introduce a completely different scheme for creating a discrete WDF.  Our approach is an extension of a discrete WDF formalism introduced by Wootters \cite{Wootters} that is constructed on the basis of a finite-dimensional quantum mechanics introduced by Schwinger \cite{Schwinger}. To apply this formalism to an open quantum dynamics system, we found that a rotationally invariant system-bath (RISB) Hamiltonian developed for the investigation of a quantum dissipative rotor system is ideal \cite{Iwamoto-2018}.  Although the bath degrees of freedom are traced out in the framework of the reduced equation of motion approach, it is important to construct a total Hamiltonian to maintain a desired symmetry of the system, including the system-bath interactions. If the symmetry of the total system is different from the main system, the quantum nature of the system dynamics is 
altered by the bath \cite{Iwamoto-2018,Iwamoto2019,Suzuki-2003}.

  Here, we employ a 2D periodically invariant system-bath (PISB) model to derive a discrete reduced equation of motion that is numerically stable regard less of the mesh size. For this purpose, we introduce two sets of the $N$-dimensional periodic operators for a momentum and coordinate spaces: The discretized reduced equation of motion is expressed in terms of these two operators,  which is stable for numerical integration even if $N$ is extremely small. The obtained equations of motion can be applied not only for a periodic system but also a system confined by a potential.  

The remainder of the paper is organized as follows. 
In Sec. 2, we introduce the periodically invariant system-bath model. In Sec. 3, we derive the HEOM for a discrete WDF. In Sec. 4, we demonstrate a stability of numerical calculations for a periodic system and a harmonic potential system using the discrete QFPE.  Sec. 5 is devoted to concluding remarks.

\section{Periodically invariant system-bath (PISB) model}
\label{sec:2}
\subsection{Hamiltonian}
We consider a periodically invariant system expressed by the Hamiltonian as
\begin{align}
\hat{H}_S &= T(\hat{p}) + U(\hat{x}),
\label{eq:system}
\end{align}
where $T(\hat{p})$ and $U(\hat{x})$ are the kinetic and potential part of the system Hamiltonian expressed as a function of the momentum and coordinate operators $\hat{p}$ and  $\hat{x}$, respectively.  In this discretization scheme, it is important that  $T(\hat{p})$ and $U(\hat{x})$ must be periodic with respect to the momentum and the coordinate, because all system operators must be written by the displacement operators in a finite-dimensional Hilbert space described subsequently.

This system is independently coupled to two heat baths through $\hat V_{x} \equiv \hbar \cos({\hat{x} dp}/{\hbar} )/dp$ and $\hat V_{y} \equiv \hbar \sin({\hat{x} dp}/{\hbar} )/dp$, where $dp$ is the mesh size of momentum \cite{Iwamoto-2018}.  Then, the PISB Hamiltonian is expressed as
\begin{align}
\hat{H}_{tot} &= \hat{H_S} + \hat V_{x} \sum_k c_k \hat{q}_{x,k} + \hat V_{y} \sum_k c_k \hat{q}_{y,k} +  \hat{H_B}, 
\end{align}
where
\begin{align}
\hat{H_B} &= \sum_k \left( \frac{\hat{p}_{x,k}^2}{2 m_k} + \frac{1}{2}m_k \omega_k \hat{q}_{x,k}^2 \right) +\sum_k \left( \frac{\hat{p}_{y,k}^2}{2 m_k} + \frac{1}{2}m_k \omega_k  \hat{q}_{y,k}^2 \right),
\label{H_B}
\end{align}
and $m_k^{\alpha}$, $\hat{p}_k^{\alpha}$, $\hat{q}_k^{\alpha}$ and $\omega_k^{\alpha}$ are the mass, momentum, position and frequency variables of the $k$th bath oscillator mode in the $\alpha = x$ or $y$ direction. The corrective coordinate of the bath $\hat{\Omega}_{\alpha}(t) \equiv \sum_k c_k\hat{q}_{\alpha,k} (t)$ is regarded as a random driving force (noise) for the system through the interactions 
${\hat{V}}_{\alpha}$. The random noise is then characterized by the canonical and symmetrized correlation functions, expressed as $\eta_{\alpha} (t) \equiv \beta \langle \hat{\Omega}_{\alpha} ; \hat{\Omega}_{\alpha}(t) \rangle_\mathrm{B}$ and $C_{\alpha}(t) \equiv \frac{1}{2} \langle \{ \hat{\Omega}_{\alpha}(t), \hat{\Omega}_{\alpha}(0) \} \rangle_{\mathrm{B}}$,
where $\beta \equiv 1/k_{\mathrm{B}}T$ is the inverse temperature divided by the  Boltzmann constant $k_\mathrm{B}$, $\hat{{\Omega}_{\alpha}}(t)$ is $\hat{{\Omega}_{\alpha}}$ in the Heisenberg representation 
and $\langle \cdots \rangle_{\mathrm{B}}$ represents the thermal average over the bath modes \cite{Tanimura89A,Tanimura-2006}.    In the classical case,  $\eta^{\alpha}(t)$  corresponds to the friction, whereas $C_{\alpha}(t)$ corresponds to the correlation function of the noise, most notably utilized in the generalized Langevin formalism.  The functions $\eta^{\alpha}(t)$ and $C_{\alpha}(t)$ satisfy the quantum version of the fluctuation-dissipation theorem, which is essential to obtain a right thermal equilibrium state \cite{Tanimura89A,Tanimura-2006,KatoJPCB13}.

The harmonic baths are characterized by the spectral distribution functions (SDF). In this paper, we assume the SDF of two heat baths are identical and are expressed as
\begin{align}
J(\omega) = \frac{\pi}{2} \sum_k \frac{(c_k)^2}{m_k \omega_k} \delta(\omega - \omega_k).
\label{Jomega} 
\end{align}
Using the spectral density $J(\omega) $, we can rewrite the friction and noise correlation function, respectively, as
\begin{align}
\eta(t) 
= \frac{2}{\pi} \int^{\infty}_0 d \omega \frac{J(\omega)}{\omega} \cos(\omega t ), 
\end{align}
and
\begin{align}
C(t)  = \frac{2}{\pi} \int^{\infty}_0 d \omega J(\omega) \coth \left(\frac{\beta\hbar\omega}{2} \right) \sin(\omega t ).
\end{align}
In order for the heat bath to be an unlimited heat source possessing an infinite heat capacity, the number of heat-bath oscillators $k$ is effectively made infinitely large by replacing $J (\omega)$ with a continuous distribution: Thus the harmonic heat baths are defined in the infinite-dimensional Hilbert space.

\subsection{System operators in a finite Hilbert space}\label{sec:2.3}
We consider a $(2 N + 1)$--dimensional Hilbert space for the system, where $N$ is an integer value.  We then introduce a discretized coordinate $X$ and momentum $P$, expressed in terms of the eigenvectors $|X, n \rangle$ and $|P, m \rangle$, where $n$ and $m$ are the integer modulo $2N + 1$ \cite{Vourdas}.  
The eigenvectors of the coordinate state satisfy the orthogonal relations
\begin{align}
 \langle X, m |X, n \rangle = \delta'_{m,n},
\end{align}
  where $\delta'_{m, n}$ is the Kronecker delta, which is equal to 1 if $n \equiv m (mod \ 2N + 1)$ (i. e. in the case that satisfies $(n-m) = (2N+1) \times$ integer), and 
\begin{align}
  \sum_{m = -N}^{N}  |X, m \rangle \langle X, m | = I,
\end{align}
where $I$ is the unit matrix.
The momentum state is defined as the Fourier transformation of the position states as
\begin{align}
|P, m \rangle = \frac{1}{2 N + 1} \sum_{n = -N}^{N} \omega^{mn} |X, n \rangle,
\end{align}
where $\omega = \exp \left[ i 2 \pi / (2N + 1) \right]$. The position and momentum operators  are then defined as
\begin{align}
\hat{x} =  \sum_{m = -N}^{N} x_m |X, m \rangle \langle X, m |, 
\end{align}
and
\begin{align}
\hat{p} =  \sum_{m = -N}^{N} p_m |P, m \rangle \langle P, m |, 
\end{align}
where $x_m = m dx$, $p_m = m dp$, and $dx$ and $dp$ are the mesh sizes of the position and momentum, respectively.  They satisfy the relation
\begin{align}
dx dp = \frac{2 \pi \hbar}{2 N + 1}.
\end{align}
To adapt the present discretization scheme, we express all system operators, including the position and momentum operators, in terms of the displacement operators (Schwinger's unitary operators \cite{Schwinger}) defined as
\begin{align}
\hat{U} _x &\equiv \exp \left(\frac{i  \hat{x}dp }{\hbar}\right), \\
\hat{U}_p &\equiv \exp \left(\frac{-i \hat{p}dx }{\hbar}\right).
\end{align}
These operators satisfy the relations, 
\begin{align}
\hat{U} _x |P, m \rangle &= |P, m + 1 \rangle, \\
\hat{U} _x |X, m \rangle &= \omega |X, m - 1 \rangle, \\
\hat{U}_p |X, m \rangle &= |X, m + 1 \rangle, \\
\hat{U}_p |X, m \rangle &= \frac{1}{\omega} |X, m - 1\rangle, 
\end{align}
and $\hat{U} _x^{2N + 1} = \hat{U} _p^{2N + 1} = I$ and $\hat{U} _x \hat{U}_p = \hat{U}_p \hat{U} _x \omega^{-1}$.
It should be noted that except in the case of $N \to \infty$, $\hat{x}$ and $\hat{p}$ do not satisfy the canonical commutation relation as in the case of the Pegg-Barnett phase operators \cite{Pegg}. (See Appendix \ref{Canonical}.)
To have a numerically stable discretization scheme, all system operators must be defined in terms of the periodic operators. Because the cosine operator in the momentum space is expanded as $\cos ( {\hat{p} dx}/{\hbar} ) = 1 - ({\hat{p} dx}/{\hbar})^2/2 +  ({\hat{p} dx}/{\hbar})^4/24 + O(dx^6)$, we defined the kinetic energy as
\begin{align}
T({\hat p}) \equiv \frac{\hbar^2}{dx^2} \left[ 1 - \cos \left( \frac{\hat{p} dx}{\hbar} \right) \right],
\label{eq:kinetic}
\end{align}
which is equivalent to $T({\hat p})   \approx \hat{p}^2/2$ with the second-order accuracy $O(dx^2)$.  As the conventional QFPE approaches use a higher-order finite difference scheme, for example a third-order \cite{SakuraiNJP14} and tenth-order central difference \cite{Ikeda2019Ohmic}, the present approach can enhance the numerical accuracy by incorporating the higher-order cosine operators, for example, as $T({\hat p})  = \hbar^2 [15 - 16\cos \left( {\hat{p} dx}/{\hbar} \right) + \cos \left( {2 \hat{p} dx}/{\hbar} \right) ]/12dx^2 + O(dx^4)$.  

Any potential $U(\hat{x})$  is also expressed in terms of the periodic operators in the coordinate space as
\begin{align}
U({\hat x}) \equiv \sum _{k = -N}^N \left[ a_k  \cos \left( k \frac{ \hat{x} dp}{\hbar} \right) +b_k \sin \left( k\frac{\hat{x} dp}{\hbar} \right)  \right],
\label{eq:pote}
\end{align}
where $a_k$ and $b_k$ are the Fourier series of the potential function $U(\hat{x})$.
The distinct feature of this scheme is that the WDF is periodic not only in the $\hat x$ space but also in the $\hat p$ space.  The periodicity in the momentum space is indeed a key feature to maintain the stability of the equation of motion. Because the present description is developed on the basis of the discretized quantum states, the classical counter part of the discrete WDF does not exist.

\section{Reduced equations of motion}
\subsection{Reduced hierarchical equations of motion}
For the above Hamiltonian with the Drude SDF
\begin{align}
J(\omega)=\frac{\eta \gamma^2 \omega}{\gamma^2+\omega^2},
\end{align}
we have the dissipation (friction)  as 
\begin{align}
\eta(t)  =\eta \exp [-\gamma t]
\end{align}
and the noise correlation functions (fluctuation) as
\begin{align}
C(t)  = c_0^{\alpha} \exp [-\gamma t] + \sum_{k=1}^{\infty} c_k^{\alpha} \exp [-k\nu t],
\end{align}
where $c_k^{\alpha}$ are the temperature dependent coefficients and $\nu=2\pi/\beta\hbar$ is the Matsubara frequency \cite{Tanimura-2006,YTperspective}.   This SDF approaches the Ohmic distribution, $J(\omega)=\eta  \omega $, for large $\gamma$. In the classical limit, both the friction and noise correlation functions become Markovian as $\eta(t) \propto \delta(t)$ and $C(t)  \propto \delta(t)$. On the other hand, in the quantum case, $C(t)$ cannot be Markovian and the value of $C(t)$ becomes negative at low temperature, owing to the contribution of the Matsubara frequency terms in the region of small $t$. This behavior is characteristic of quantum noise. The infamous positivity problem of the Markovian QME for a probability distribution of the system arises due to the unphysical Markovian assumption under the fully quantum condition \cite{Tanimura-2006,YTperspective,YTJCP2014}. The fact that the noise correlation takes negative values introduces problems when the conventional QFPE is applied to quantum tunneling at low temperatures \cite{YTJCP2015}.

Because the HEOM formalism treats the contribution from the Matsubara terms accurately utilizing hierarchical reduced density operators in a non-perturbative manner, there is no limitation to compute the dynamics described by the system-bath Hamiltonian \cite{TanimuraPRA91,TanimuraJCP92,Tanimura-2006,YTperspective,YTJCP2014,YTJCP2015,SakuraiJPSJ13,SakuraiNJP14,GrossmanSakuraiJPSJ,KatoJPCB13,MS-QFP94,MS-QFPChernyak96,TanimuraMaruyama97,MaruyamaTanimura98,Ikeda2019Ohmic,Ikeda2017JCP}.
The HEOM for the 2D PISB model is easily obtained from those for the three-dimensional RISB model as \cite{Iwamoto2019} 
\begin{align}
\frac{\partial }
{{\partial t}}\hat \rho _{\{ n_{\alpha} \}} (t) = & - \left( {\frac{\rm i}
{\hbar}\hat H_S^ \times   + \sum_{\alpha=x,y}  n_{\alpha}\gamma } \right)\hat \rho_{\{ n_{\alpha} \}} (\,t) \nonumber \\
&- \sum_{\alpha=x,y}  \frac{\rm i}
{\hbar}\hat V_{\alpha}^ \times  \hat \rho _{\{ n_{\alpha}+1 \}} (t) - \sum_{\alpha=x,y}  \frac{{{\rm i}n_{\alpha}}}
{\hbar }\hat \Theta_{\alpha} \hat \rho _{\{ n_{\alpha}-1 \}} (t), 
\label{eq:GMmaster}
\end{align}
where $\{ n_{\alpha} \}\equiv (n_x, n_y)$  is a set of integers to describe the hierarchy elements and $\{n_{\alpha}\pm1 \}$ represents, for example, $ (n_x, n_y\pm 1)$ for $\alpha=y$, and
\begin{eqnarray}
\hat \Theta_{\alpha}  \equiv \eta \gamma
\left(\frac{ 1}{\beta}{\hat V_{\alpha}^ \times   - \frac{ \hbar}{2 }\hat H_S^ \times \hat V_{\alpha}^\circ  } \right),
\label{eq:Theta_S}
\end{eqnarray}
with $\hat A^\times \hat \rho \equiv \hat A \hat \rho - \hat \rho \hat A$ and $\hat A^\circ \hat \rho \equiv \hat A \hat \rho + \hat \rho \hat A$ for any operator $\hat A$. We set $\hat \rho _{\{n_{\alpha}-1\}} (t)=0$ for $n_{\alpha}=0$. 

For $(n_{\alpha} + 1) \gamma \gg \eta /\beta$ and $(n_{\alpha} + 1) \gamma \gg \omega_0$ (the high temperature Markovian limit), where $\omega_0$ is the characteristic frequency of the system, we can set $ {\rm i}\hat V_{\alpha}^ \times \hat \rho _{\{ n_{\alpha}+1 \}} (t)/\hbar = \hat \Gamma_{\alpha} \hat \rho _{\{ n_{\alpha} \}} (t) $ to truncate the hierarchy, where 
\begin{eqnarray}
\hat \Gamma_{\alpha} \equiv  \frac{{ 1 }}{{\gamma \hbar^2 }}\hat V_{\alpha}^ \times  \hat \Theta_{\alpha}
\label{eq:Gamma}
\end{eqnarray}
is the damping operator \cite{Tanimura89A,TanimuraPRA91,TanimuraJCP92,Tanimura-2006,YTperspective,YTJCP2014,YTJCP2015}.

In a high temperature Markovian case with $J(\omega) = \eta \omega/\pi$, the HEOM reduces to the Markovian QME without the rotating wave approximation (RWA) expressed as \cite{Iwamoto-2018}
\begin{align}
\frac{\partial}{\partial t} \hat{\rho}(t)  &=\frac{\rm i}
{\hbar}{\hat H}_S^ \times \hat{\rho}(t) 
- \frac{1}{\beta \hbar^2} \sum_{\alpha=x,y}  \hat \Gamma_{\alpha}  \hat{\rho}(t).
\label{eq:QMEiso}
\end{align}
To demonstrate a role of the counter term in the present 2D PISB model, we derive the above equation from the perturbative approach in Appendix \ref{QMEcounter}.

\subsection{Discrete quantum hierarchical Fokker-Planck equation}

The HEOM for the conventional WDF have been used for the investigation of various problems \cite{TanimuraPRA91,TanimuraJCP92,Tanimura-2006,YTperspective,YTJCP2014,YTJCP2015,KatoJPCB13,TanimuraMaruyama97,MaruyamaTanimura98,Ikeda2019Ohmic,Ikeda2017JCP}, 
including the RTD problem \cite{SakuraiJPSJ13,SakuraiNJP14,GrossmanSakuraiJPSJ}. Here we introduce a different expression on the basis of a discrete WDF.  While there are several definitions of a discrete WDF \cite{Galetti,luis,klimov}, in this paper, we use a simple expression introduced by Vourdas \cite{Vourdas}. For any operator $\hat{A}$ is then expressed in the matrix form as
\begin{align}
\bm{A}(p_j, q_k) &= \sum_{l = -N}^{N} \exp \left(i \frac{2 p_j (q_k - q_l)}{\hbar} \right) \langle X, l | \hat{A} | X, 2 k - l \rangle \\
&= \sum_{l = -N}^{N} \exp \left(i \frac{-2 q_k (p_j - p_l)}{\hbar} \right) \langle P, l | \hat{A} | P, 2 j - l \rangle,
\end{align}
where we introduced $q_k = k dx$ and $p_j = j dp$.  For $\hat{A}=\hat{\rho}$, we have the discrete WDF expressed as \bm{$W$}$(p_j, q_k)$. 
This discrete WDF is analogous to the conventional WDF, although the discretized regions in the $p$ and $q$ spaces are both from -$N$ to $N$ and are periodic in this case. 
Thus, for example, for $k<-N$, we have   $k \to k + 2N+1$ and $q_k = (k+2N+1)dx$, and for $j>N$, we have $j \to j - 2N-1$ and $p_j = (j-2N-1)dp$.  The Wigner representation of the reduced equations of motion, for example Eq. \eqref{eq:GMmaster} and Eq.\eqref{eq:QMEiso} can be obtained by replacing the product of any operators $\hat{A}_1$ and $\hat{A}_2$ by the star product defined as
\begin{align}
[\bm{A_1} \star \bm{A_2}](p_j, q_k) &\equiv \frac{1}{(2 N + 1)^2} \sum_{j_1, j_2, k_1, k_2 = -N}^N \exp \left( i \frac{2 p_{j_2} q_{k_1} - 2 p_{j_1} q_{k_2}}{\hbar} \right) \notag \\
&\times  \bm{A_1} (p_j + p_{j_1}, q_k + q_{k_1})  \bm{A_2}( p_j + p_{j_2}, q_k + q_{k_2}).
\label{eq:Moyal}
\end{align}
Accordingly, the quantum commutator $[ \ ,\ ]$ is replaced as the discrete Moyal bracket defined as $\{ \bm{A_1} , \bm{A_2} \}_M \equiv \bm{A_1} \star \bm{A_2} - \bm{A_2} \star \bm{A_1} $.

The HEOM in the desecrate WDF (the discrete QHFPE) are then expressed as
\begin{align}
\frac{\partial}{\partial t} \bm{W} _{\{ n_{\alpha} \}}  &= -\frac{i}{\hbar} \{ \bm{H_S} , \bm{W} _{\{ n_{\alpha} \}} \}_M \notag \\
& + \sum_{\alpha=x,y}  n_{\alpha}\gamma  \bm{W} _{\{ n_{\alpha} \}} - \sum_{\alpha=x,y}  \frac{\rm i}{\hbar} \{  \bm{V_{\alpha}} ,  \bm{W} _{\{ n_{\alpha}+1 \}} (t) \}_M \notag \\
&- \sum_{\alpha=x,y}  \frac{{{\rm i}n_{\alpha} \eta \gamma}}
{\beta \hbar } \left( \{  \bm{V_{\alpha}} , \bm{W} _{\{ n_{\alpha}-1 \}} (t)  \}_M \right. \notag \\
&\left. - \frac{ \hbar}{2 } \{ \bm{H_S} , \bm{\hat V_{\alpha}} \star \bm{W} _{\{ n_{\alpha}-1 \}} (t) + \bm{W} _{\{ n_{\alpha}-1 \}} (t) \star \bm{\hat V_{\alpha}}   \}_M  \right) .
\label{eq:HEOMWigner}
\end{align}
As illustrated by Schwinger \cite{Schwinger}, although we employed the periodic WDF, we can investigate the dynamics of a system confined by a potential by taking the limit $N \to \infty $ for $dx = x_0 \sqrt{2 \pi/(2N + 1)}$  and $dp = p_0 \sqrt{2 \pi/(2N + 1)}$ with $x_0 p_0 = \hbar$, while we set $dx = L/(2N + 1)$ and $dp = 2 \pi \hbar/L$ in the periodic case, where $L$ is the periodic length.

\subsection{Discrete quantum Fokker-Planck equation}

In the high temperature Markovian limit, as the regular HEOM (Eq. \eqref{eq:GMmaster}) reduces to the QME (Eq.\eqref{eq:QMEiso}), the discrete QHFPE reduces to the discrete QFPE expressed as 
\begin{align}
\frac{\partial}{\partial t} \bm{W}  &= -\frac{i}{\hbar} \{ \bm{H_S} , \bm{W} \}_M \notag \\
& + \sum_{\alpha = x, y} \left[ -\frac{\eta}{\beta \hbar^2}  \left(  \{\bm{\hat{V}_{\alpha}},  \bm{\hat{V}_{\alpha}} \star \bm{W}\} _M- \{ \bm{\hat{V}_{\alpha}} ,   \bm{W} \star \bm{\hat{V}_{\alpha}}\}_M  \right) \right. \notag \\
& + \frac{\eta}{2 \hbar^2} \left( \{\bm{\hat{V}_{\alpha}} ,  \bm{H_S} \star \bm{\hat{V}_{\alpha}} \star \bm{W} \}_M   + \{ \bm{\hat{V}_{\alpha}} ,   \bm{H_S}  \star \bm{W} \star \bm{\hat{V}_{\alpha}} \}_M   \right. \notag \\
&\left. \left. - \{\bm{\hat{V}_{\alpha}} ,  \bm{\hat{V}_{\alpha}} \star \bm{W}  \star \bm{H_S}\}_M   - \{ \bm{\hat{V}_{\alpha}} ,   \bm{W} \star \bm{\hat{V}_{\alpha}} \star  \bm{H_S}\}_M  \right) \right].
\label{eq:QMEWigner}
\end{align}
Here, the terms proportional to $\eta/\beta \hbar^2$ and $\eta/2 \hbar^2$ represent the effects of thermal fluctuation and dissipation arise from the heat bath, respectively. 
More explicitly, the above equation is expressed as (Appendix \ref{MoyalBracket})
\begin{align}
\frac{\partial}{\partial t} W(p_j, q_k) &= -\hbar \sin \left(\frac{p_j dx }{\hbar} \right) \frac{W(p_j, q_{k + N + 1}) - W(p_j, q_{k - N - 1} )}{dx^2} \notag \\
&-\frac{i}{\hbar} \{ \bm{U} , \bm{W} \}_M \notag \\
&+ \frac{\eta}{ \beta} \frac{W(p_{j + 1}, q_k) - 2 W(p_j, q_k)+ W(p_{j - 1}, q_k)}{dp^2} \notag \\
&-\frac{\hbar^2 \eta (\omega - 2 + \omega^{-1}) V_{p_j}}{4 dx^2 dp^2}\left( W(p_j, q_{k + N + 1}) + W(p_j, q_{k - N - 1}) \right) \notag \\
&+ \frac{\hbar^2 \eta (V_{p_j} - V_{p_{j+1}})}{4 dx^2 dp^2}(W(p_{j + 1} , q_{k + N + 1}) + W(p_{j + 1} , q_{k - N - 1}))  \notag \\
&+ 
\frac{\hbar^2 \eta (V_{p_j} - V_{p_{j-1}})}{4 dx^2 dp^2}
(W(p_{j - 1} , q_{k + N + 1}) + W(p_{j - 1} , q_{k - N - 1})),
\label{eq:QMEWignerDIS}
\end{align}
where $V_{p_j} \equiv \cos({p_j dx }/{\hbar})$.
The numerical stability of the above equation arises from the finite difference scheme in the periodic phase space.  For example, the finite difference of the kinetic term (the first term in the RHS of Eq.\eqref{eq:QMEWignerDIS}) is constructed from the elements not the vicinity of $q_k$ (i.e. $q_{k +1}$ and $q_{k-1}$), but the boundary of the periodic $q$ state (i.e. $q_{k + N + 1}$ and $q_{k - N - 1}$).  The dissipation terms (the last three terms in the RHS of Eq.\eqref{eq:QMEWignerDIS}) are also described by the boundary elements.  As we will show the harmonic case below, the potential term (the second term in the RHS of Eq.\eqref{eq:QMEWignerDIS}) is constructed from the boundary of the periodic $p$ state.  Because Eq. \eqref{eq:QMEWignerDIS} satisfies $\sum_{k = -N}^N \sum_{j = -N}^N W(p_k,q_j)=1$  and because the operators in the discrete QFPE are non-local, the calculated results are numerically stable regardless of a mesh size. 

For large $N$, we have $\sin ({p_j dx }/{\hbar} ) \approx p_j dx /{\hbar}$ and  $\cos({p_j dx }/{\hbar}) \approx 1 -( p_j dx /{\hbar})^2$.  Then the above equation is expressed in a similar form as the QFPE  obtained by Caldeira and Leggett \cite{CaldeiraPhysica83,TanimuraPRA91}, although the finite difference expressions for discrete WDF are quite different from those for the conventional WDF. (See Appendix \ref{QFPELargeN}).

\section{Numerical results}

In principle, with the discrete WDF, we are able to compute various physical quantities by adjusting the mesh size determined from $N$ for any periodic system and a system confined by a potential.
A significant aspect of this approach is that even small $N$, the equation of motion is numerically stable, although accuracy may not be sufficient.  

In the following, we demonstrate this aspect by numerically integrating Eq. \eqref{eq:QMEWignerDIS} for the a free rotor case and a harmonic potential case, for which we have investigated from the regular QME approach  \cite{Iwamoto-2018} and QFPE approach  \cite{TanimuraPRA91}. 
In both cases, we considered a weak damping condition ($\eta=0.05$) at high temperature ($\beta=0.1$).  For time integrations, we used the fourth-order Runge-Kutta method with the step $\delta t=0.001$. 
In the free rotor case, we chose $N$ to minimize the momentum space distribution near the boundary, whereas, in the harmonic case, we chose $N$ 
to minimize the population of the discrete WDF near the boundary in both the $q$ and $p$ directions.

\begin{figure}
\centering
\includegraphics[width=0.9\textwidth]{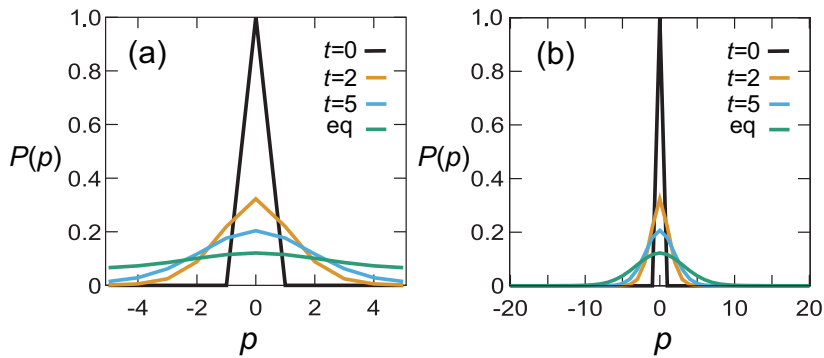}
\caption{Snapshots of the momentum space distribution function, $P(p_k) = \sum_{j = -N}^N W(p_k,q_j)$, in the free rotor case calculated from Eq. \eqref{eq:QMEWignerDIS} for mesh size (a) $N=5$ and (b) $N=20$ with the waiting times $t=0, 2, 5$, and $100$ (equilibrium state). 
\label{freePp}
}
\end{figure}

\begin{figure}
\centering
\includegraphics[width=0.9\textwidth]{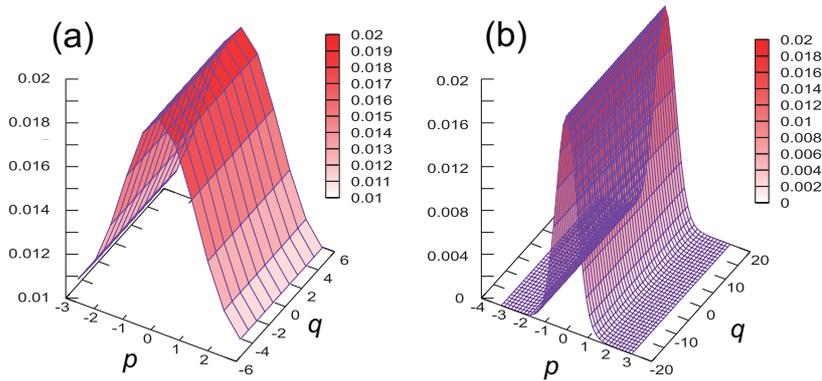}
\caption{The equilibrium distribution ($t=100$) of the discrete WDF in the free rotor case for (a) $N=5$ and (b) $N=20$.
\label{eqWDF}
}
\end{figure}

\subsection{Free rotor case}
We first examine the numerical stability of Eq.~\eqref{eq:QMEWignerDIS} for a simple free rotor case,  $U(\hat{x})= 0$ with $L = 2 \pi$.  For demonstration, we considered localized initial conditions expressed as $W(p_0, q_j)=1$ for $-N \le j \le N$ with $p_0=0$, and  zero otherwise. 
While such initial conditions that involve a singularity in the $p$ direction is not easy to conduct numerical simulation from a conventional finite difference approach, there is no difficulty from this approach.  Moreover, the total population is alway conserved within the precision limit of the numerical integration, because we have $\sum_{k = -N}^N \sum_{j = -N}^N W(p_k,q_j)=1$. 

We first depict the time evolution of the momentum distribution function $P(p_k) = \sum_{j = -N}^N W(p_k,q_j)$  for (a)  $N = 5$ and (b) $20$, respectively.  Here, we do not plot $P(q_j) = \sum_{k= -N}^N W(p_k, q_j)$, because this is always constant as a function of $q$, as expected for the free rotor system.
As illustrated in Fig. \ref{freePp}, even the distribution was localized at $p_0=0$ at $t=0$, calculated $P(p_k)$ was alway stable.  As the waiting time increased, the distribution became a Gaussian-like profile in the $p$ direction owing to the thermal fluctuation and dissipation both of which arose from the heat bath.
In this calculation, the larger $N$ we used, the more accurate results we had.  We found that the results converged approximately $N=20$, and coincided with the results obtained from the conventional QME approach with use of the finite difference scheme \cite{Iwamoto-2018}.  The equilibrium distributions of the discrete WDF  for different $N$ are depicted in Fig. \ref{eqWDF}. As $N$ increases, the distribution in the $p$ direction approached the Gaussian profile.

\begin{figure}
\centering
\includegraphics[width=0.5\textwidth]{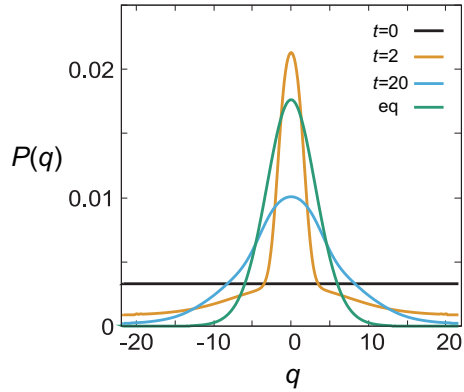}
\caption{Snapshots of the coordinate space distribution function, $P(q_j) = \sum_{k= -N}^N W(p_k, q_j)$, in the harmonic potential case calculated from Eq. \eqref{eq:QMEWignerDIS} with $N=150$ with the waiting times $t=0, 2, 20$, and $200$ (equilibrium state). 
\label{HoPx}
}
\end{figure}

\begin{figure}
\centering
\includegraphics[width=0.5\textwidth]{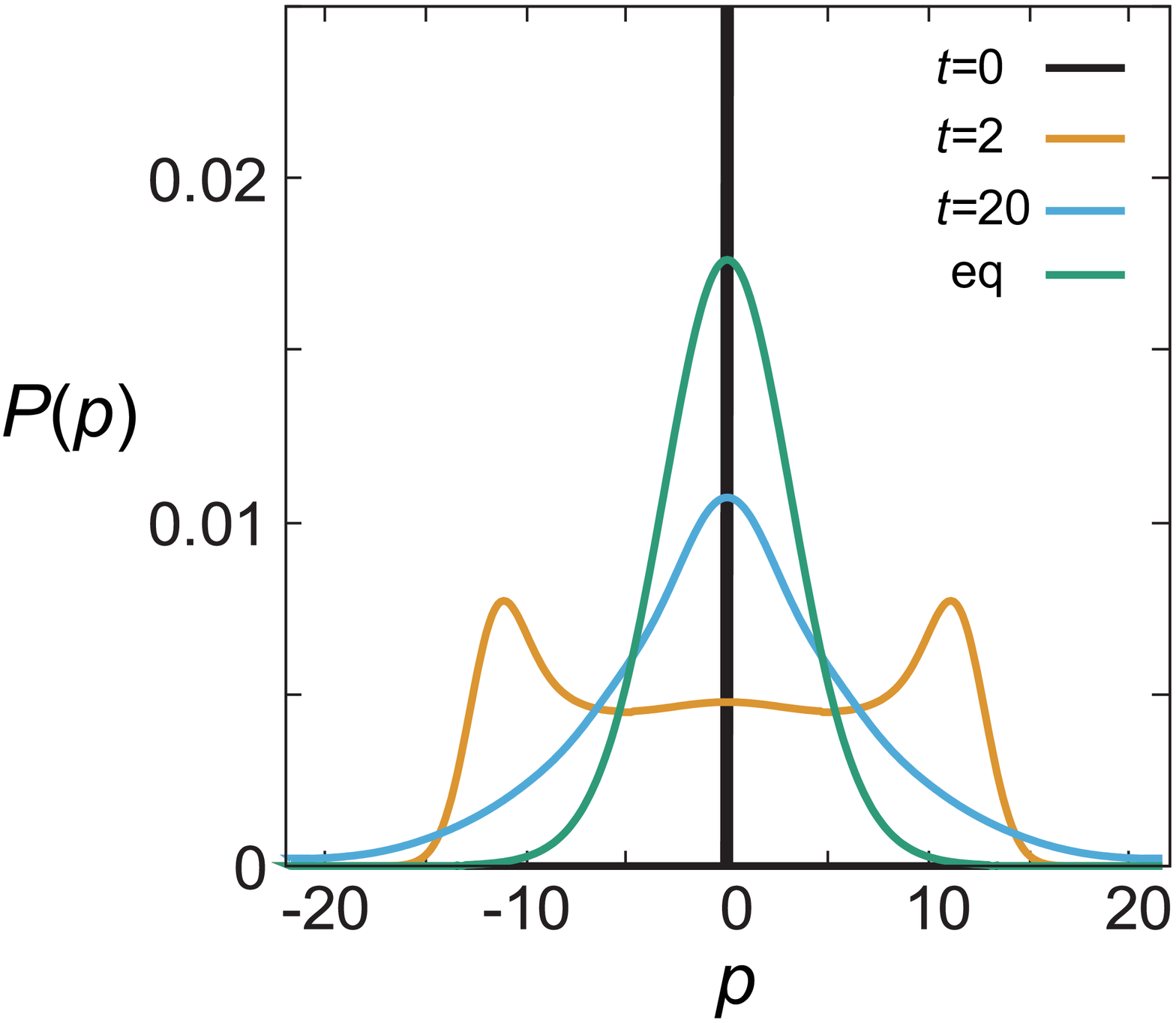}
\caption{Snapshots of the momentum space distribution function, $P(p_k) = \sum_{j = -N}^N W(p_k,q_j)$, in the harmonic potential case calculated from Eq. \eqref{eq:QMEWignerDIS} with $N=150$ with the waiting times $t=0, 2, 20$, and $200$ (equilibrium state). 
\label{HoPp}
}
\end{figure}

\begin{figure}
\centering
\includegraphics[width=0.9\textwidth]{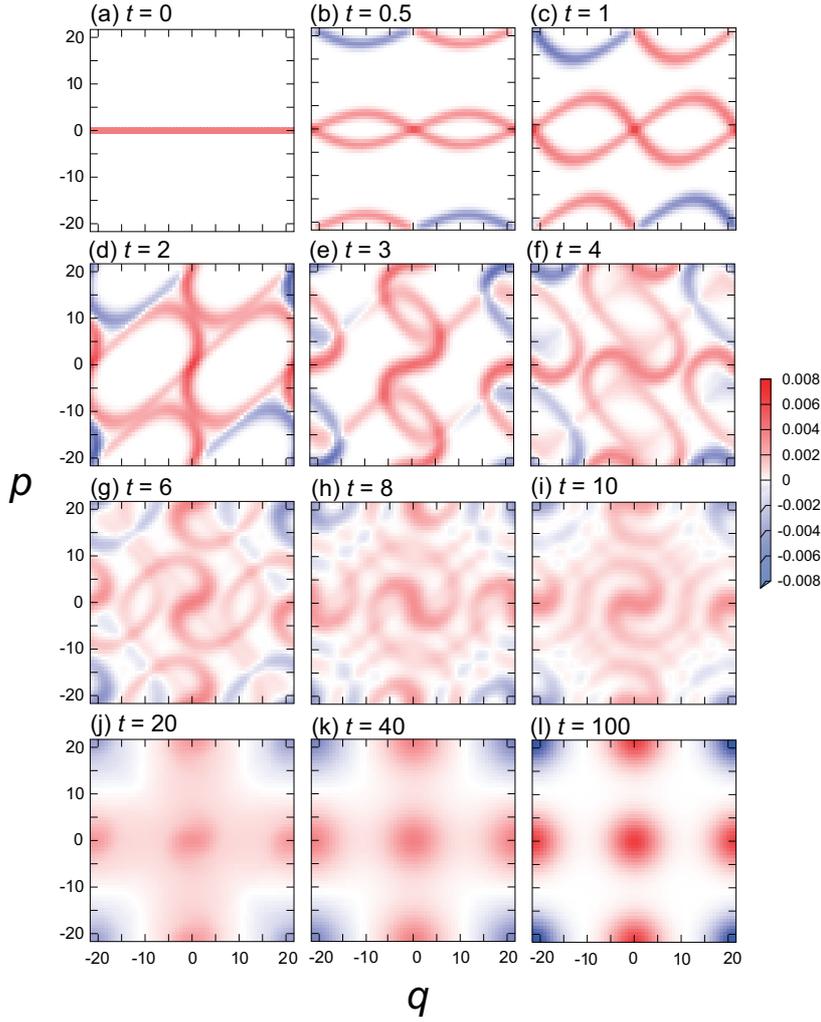}
\caption{Snapshots of  the discrete WDFs in the harmonic case for various values of the waiting time. Contours in red and blue represent positive and negative values, respectively. The mesh size is $N=150$.
\label{eqHoWDF}
}
\end{figure}

\subsection{Harmonic case}
We next consider a harmonic potential case, $U(\hat{x})=\hat{x}^2/2$.
Here, we describe the potential using the periodic operator as $U(\hat{x}) \approx \hbar^2 (1 - \cos(
 {\hat{x} dp}/{\hbar} ) )/dp^2 + O(dp^2)$. Then the potential term is expressed as
\begin{align}
-\frac{i}{\hbar} \{ \bm{U} , \bm{W} \}_M = \hbar \sin \left(\frac{x_k dp}{\hbar} \right) \frac{W(p_{j + N + 1}, q_k) - W(p_{j - N - 1}, q_k)}{dp^2}.
\label{eq:potential}
\end{align}
With this expression, we simulated the time-evolution of the discrete WDF by numerically  integrating Eq.\eqref{eq:QMEWignerDIS} with $N = 150$. We chose the same system-bath coupling strength and inverse temperature as in the free rotor case (i.e. $\eta=0.05$ and $\beta=0.1$).  In Figs. \ref{HoPx} and \ref{HoPp}, we depict the time-evolution of the position and momentum distribution functions $P(q_j) $ and  $P(p_k) $ in the harmonic case from the same localized initial conditions as in the free rotor case.
As illustrated in Figs. \ref{HoPx} and \ref{HoPp}, both $P (q_j)$ and $P(p_k)$ approached  the Gaussian-like profiles as analytic derived solution of the Brownian model predicted.  Note that, although the discrete WDF is a periodic function, we can describe such distribution that is confined in a potential by combining the periodicity  in the coordinate  and momentum spaces.  

To illustrate a role of periodicity, we depict the time-evolution of the discrete WDF for various values of the waiting time in Fig. \ref{eqHoWDF}.   At time (a) $t=0$, the distribution was localized at $p_0=0$, while the distribution in the $q$ direction was constantly spread.   At time (b) $t=0.5$, the distribution symmetrically splits into the positive and negative $p$ directions because the total momentum of the system is zero.  Due to the kinetic operator (the first term in the RHS in Eq.\eqref{eq:QMEWignerDIS}), the vicinity of the distributions at $(p, q )=(p_0, q_{-(N+1)} )$, and $(p_0, q_{N+1})$ appeared and the profiles of the distribution are similar to the distribution near $(p_0, q_0)$.  
We also observed the distribution along the vicinity of the $p=p_{N+1}$ and $p=p_{-(N+1)}$ axises, respectively.  These distributions arose owing to the finite difference operator of the potential term in Eq. \eqref{eq:potential}, which created the positive and negative populations $W(p_{N + 1}, q_k)$ and $-W(p_{ - (N+1)}, q_k)$ from $W(p_0, q_k)$.  The sign of these distributions changed at $q=q_0$, because of the presence of the prefactor $\sin \left({x_k dp}/{\hbar} \right)$. 
In Fig. \ref{eqHoWDF}(d), we observed the tilted $x$ letter-like distributions centered at  $(p, q)=(12, 0)$ and $(-12, 0)$, respectively. These distributions appeared as twin peaks in the momentum distribution depicted in Fig. \ref{HoPp}. 

From time (e) $t=1.0$ to (i) $t=10$, the distributions rotated clockwise to the centered at $(q_0, p_0)$ as in the conventional WDF case.  
Owing to the periodic nature of the kinetic and potential operators, we also observe the mirror images of the central distribution at $(q_0, p_{N + 1})$, $(q_0, p_{-(N+1)})$,  $(q_{-(N+1)}, p_0 )$, and $(q_{N+1}, p_0)$, respectively.  At time (j) $t=20$, the spiral  structures of the distributions disappeared owing to the dissipation, then the profiles of distributions became circular.  The peaks of the circular distributions became gradually higher due to the thermal fluctuation (excitation), as depicted in Fig. \ref{eqHoWDF} (k) $t=40$. The distributions were then reached to the equilibrium profiles,  in which the energy supplied by fluctuations and the energy lost through dissipation were balanced, as presented in Fig. \ref{eqHoWDF} (i) $t=100$.  It should be noted that, although $P (q_j)$ and $P (p_k)$  in  Figs. \ref{HoPx} and \ref{HoPp} exhibited the Gaussian profiles, each circular distribution observed in Fig. \ref{eqHoWDF} (l) need not be the Gaussian, because the discrete WDF itself is not a physical observable. The negative distributions in the four edges of the phase space arose due to the prefactors of the kinetic and potential terms  $\sin \left({x_k dp}/{\hbar} \right)$ and  $\sin \left({p_j dx }/{\hbar} \right)$.   Although the appearance of the discrete WDF is very different from the conventional WDF, this is not surprising, because the discrete WDF does not have classical counter part.  This unique profile of the discrete WDF is a key feature to maintain the numerical stability of the discrete QFPE.

\section{Conclusion}

In this paper, we developed an open quantum dynamics theory for the discrete WDF. Our approach is based on the PISB model with a discretized operator defined in the $2N+1$ periodic eigenstates  in both the $q$ and $p$ spaces.  The kinetic, potential, and system-bath interaction operators  in the
equations of motion are then expressed in terms of the periodic operators; it provide numerically stable discretization scheme regardless of a mesh size.  The obtained equations are applicable not only for a periodic system but also a  system confined by a potential. We demonstrated the stability of this approach in a Markovian case by integrating the discrete QFPE for a free rotor and harmonic cases started from singular initial conditions.  
 It should be noted that the Markovian condition can be realized only under high-temperature conditions even if we consider the Ohmic SDF due to the quantum nature of the noise.  To investigate a system in a low temperature environment, where quantum effects play an essential role, we must  include low-temperature correction terms in the framework of the HEOM formalism \cite{Tanimura-2006,YTperspective}, for example the QHFPE  \cite{YTJCP2014} or the low-temperature corrected QFPE \cite{Ikeda2019Ohmic}.

As we numerically demonstrated,  we can reduce the computational cost of dynamics simulation by suppressing the mesh size, while we have to examine the accuracy of the results carefully. 
 If necessary, we can employ a present model with small $N$ as a phenomenological model for an investigation of a system described by a multi-electronic and multi-dimensional potential energy surfaces, for example, an open quantum system that involves a conical intersection \cite{Ikeda2018CI}.

Finally, we briefly discuss some extensions of the present study. In the current frameworks, it is not easy to introduce open boundary conditions, most notably the in flow and out flow boundary conditions \cite{Ringhofer1989,Jiang2014,Mahmood2016},  because our approach is constructed on the basis of the periodical phase space.
 Moreover, when the system is periodic, it is not clear whether we can include a non-periodical external field, for example a bias field \cite{SakuraiJPSJ13,SakuraiNJP14,GrossmanSakuraiJPSJ} or ratchet refrigerant forces \cite{KatoJPCB13}. Moreover, numerical demonstration of the discrete QHFPE (Eq. \eqref{eq:HEOMWigner}) has to be conducted for a strong system-bath coupling case at low temperature.  Such extensions are left for future investigation.

\section*{Acknowledgements}
Y.~T.~is supported by JSPS KAKENHI Grant Number B 21H01884.

\section*{Conflict of interest}
The authors declare that they have no conflict of interest.

\section*{Data availability}
The data that support the findings of this study are available from the corresponding author upon a reasonable request.

\appendix
\section{Canonical commutation relation in the large $N$ limit}
\label{Canonical}
In this Appendix, we show that our coordinate and momentum operators satisfy the canonical commutation relation in the large $N$ limit.

First we consider a non-periodic case, $dx = x_0 \sqrt{2 \pi/(2N + 1)}$  and $dp = p_0 \sqrt{2 \pi/(2N + 1)}$ with $x_0 p_0 = \hbar$. We employ the relationship between the displaced operator, $\hat{U}_x \hat{U}_p - \hat{U}_p \hat{U}_x \omega^{-1} = 0$. 
Assuming large $N$, we express $\hat{U}_x$ and $\hat{U}_p$ in Taylor expansion forms as
\begin{align}
&\left[1 + \frac{i dp \hat{x}}{\hbar} + \frac{(i dp)^2}{2 \hbar^2} \hat{x}^2\right]\left[1 + \frac{i \hat{p} dx}{\hbar} + \frac{(i dx)^2}{2 \hbar^2} \hat{p}^2 \right]  \notag \\
&- \left[1 + \frac{i \hat{p} dx}{\hbar} + \frac{(i dx)^2}{2 \hbar^2} \hat{p}^2\right] \left[1 + \frac{i dp \hat{x}}{\hbar} + \frac{(i dp)^2}{2 \hbar^2} \hat{x}^2\right] \left[1 - \frac{i dx dp}{\hbar}) + O( (N^{\frac{-3}{2}})\right] \nonumber \\
&= \frac{dx dp}{\hbar^2}(\hat{x} \hat{p} - \hat{p} \hat{x}) - \frac{i dx dp}{\hbar} + O( N^{\frac{-3}{2}} ).
\end{align}
This indicates that the canonical commutation relation  $[\hat{x} , \hat{p}] = i \hbar$ satisfies to an accuracy of $O( N^{\frac{-3}{2}} )$.

In the $2 \pi$-periodic case, we set $dx = {2 \pi}/{(2 N + 1)}$ and  $dp = \hbar$.
Then we obtain 
\begin{align}
&(\cos \hat{x} + i \sin \hat{x}) \left(1 - \frac{i dx}{\hbar} \hat{p} \right) - \left(1 - \frac{i dx}{\hbar} \hat{p}\right) (\cos \hat{x} + i \sin \hat{x}) (1 - i dx) + O(N^{-2}) \nonumber \\
&= \frac{dx}{\hbar}(\sin \hat{x} \hat{p} - \hat{p} \sin \hat{x} - i \hbar \cos \hat{x}) + \frac{i dx}{\hbar}(\cos \hat{x} \hat{p} - \hat{p} \cos \hat{x} + i \hbar \sin \hat{x}) + O(N^{-2}).
\label{eq:CCR} 
\end{align}
The first and second terms of the RHS in Eq. \eqref{eq:CCR} are the anti-Hermite and Hermite operators. Therefore, the contributions from these terms are zero. 
Thus, for large $N$, we obtain the canonical commutation relations for a periodic case as \cite{Curruthers}
\begin{align}
[\sin \hat{x} , \hat{p}] &= i \hbar \cos \hat{x},
\end{align}
and
\begin{align}
[\cos \hat{x} , \hat{p}] &= - i \hbar \sin \hat{x},
\end{align}
to an accuracy of $ O(N^{-2})$.

\section{QME for 2D PISB model and counter term}
\label{QMEcounter}
To demonstrate a role of the counter term, here we employ the QME for the 2D PISB model.  As shown in \cite{Iwamoto-2018},
 the QME for the reduced density matrix of the system, $\hat{\rho}(t)$, is derived from the second-order perturbation approach as  
\begin{align}
\frac{\partial}{\partial t} \hat{\rho}(t)  &= -\frac{i}{\hbar}[\hat{H}_S , \hat{\rho}(t)] - \frac{1}{\hbar^2} \int^{t}_{0} d \tau \left(\hat \Gamma_x(\tau) \hat{\rho}(t - \tau) + \hat \Gamma_y(\tau) \hat{\rho}(t - \tau)\right),
\label{eq:GQME}
\end{align}
where 
\begin{align}
\hat \Gamma_\alpha(\tau) \hat{\rho}(t - \tau) &\equiv C(\tau) [\hat{V}_{\alpha} , \hat{G}_S(\tau) \hat{V}_{\alpha} \hat{\rho}(t - \tau) \hat{G}_S^{\dagger}(\tau)] \notag \\
& - C(-\tau)[ \hat{V}_{\alpha} ,  \hat{G}_S(\tau) \hat{\rho}(t - \tau) \hat{V}_{\alpha} \hat{G}_S^{\dagger}(\tau)]
\label{eq:Gamma}
\end{align}
is the damping operator for $\alpha = x$ or $y$, in which
\begin{align}
C(\tau) &= \hbar \int^{\infty}_0 \frac{d \omega}{\pi} J (\omega) \left[ \coth \left(\frac{\beta \hbar \omega}{2}\right) \cos(\omega \tau) - i \sin(\omega \tau)\right]
\label{eq:spectral}
\end{align}
is the bath correlation function and $\hat{G}_S(\tau)$ is the time evolution operator of the system. For the Ohmic SDF $J(\omega)=\eta \omega $, $C(\tau)$ reduces to the Markovian form as
 \begin{align}
C(\tau) = \eta \left( \frac{2 }{\beta}  + i
 \hbar  \frac{d}{d\tau} \right) \delta(\tau).
\end{align}
Using the relation $\int^{t}_{0} d \tau \hat \Gamma_{\alpha}(\tau) \hat{\rho}(t - \tau) = \hat {\bar \Gamma}_{\alpha} \hat{\rho}(t)+{i \hbar \eta }\delta(0) [(\hat{V}_{\alpha})^2, \hat{\rho}(t)]$, we can rewrite the damping operator, Eq. \eqref{eq:Gamma}, as 
\begin{align}
\hat {\bar \Gamma}_{\alpha}  \hat{\rho}(t)&= 
 \frac{\eta}{\beta}  \left(  [\hat{V}_{\alpha},  \hat{V}_{\alpha} \hat{\rho}(t) ] - [ \hat{V}_{\alpha} ,   \hat{\rho}(t ) \hat{V}_{\alpha}]  
\right) +\frac{i\hbar \eta}{2 } \left[(\hat{V}_{\alpha})^2 ,  \frac{d \hat{\rho}(t - \tau)}{d \tau}|_{\tau = 0} \right]\notag \\
&- \frac{\eta}{2 } \left([\hat{V}_{\alpha} ,  \hat{H}_S \hat{V}_{\alpha} \hat{\rho}(t ) ] + [ \hat{V}_{\alpha} ,   \hat{H}_S \hat{\rho}(t ) \hat{V}_{\alpha} ] - [\hat{V}_{\alpha} ,  \hat{V}_{\alpha} \hat{\rho}(t ) \hat{H}_S] - [ \hat{V}_{\alpha} ,   \hat{\rho}(t ) \hat{V}_{\alpha} \hat{H}_S]  \right).
\label{eq:anisoQME}
\end{align}
In the case if there is only $\hat V_{y} =\hbar \sin({\hat{x} dp}/{\hbar} )/dp$ interaction in the PISB model, (i.e. $\hat{V}_{x}=0$), we encounter the divergent term ${i \hbar \eta }\delta(0) [(\hat{V}_{y})^2, \hat{\rho}(t)]$ that arises from the second term in the RHS of Eq. \eqref{eq:anisoQME}.  Because $\hat{V}_{y}$ reduces to the linear operator of the coordinate $\hat{V}_{y} \approx \hat x$ in the large $N$ limit, the PISB model under this condition corresponds to the Caldeira-Leggett model without the counter term:  Divergent term arises because we exclude the counter term in the bath Hamiltonian, Eq. \eqref{H_B}.  (See also  \cite{Suzuki-2003}.)
If we include $\hat V_{x} =\hbar \cos({\hat{x} dp}/{\hbar} )/dp$, this divergent term vanishes, because, by using the relation $\sin^2({\hat{x} dp}/{\hbar} ) + \cos^2({\hat{x} dp}/{\hbar} )=1$, we have 
\begin{align}
{i \hbar \eta }\delta(0) [(\hat{V}_{x})^2, \hat{\rho}(t)] + {i \hbar \eta }\delta(0) [(\hat{V}_{y})^2, \hat{\rho}(t)] &= {i \hbar \eta }\delta(0)[\hat{I}, \hat{\rho}(t)] \notag \\
&= 0.
\end{align}
This implies that the interaction $\hat{V}_{y}$ plays the same role as the counter term.  This fact indicates the significance of constructing a system-bath model with keeping the same symmetry as the system itself. If we ignore this point, the system dynamics are seriously altered by the bath even if the system-bath interaction is feeble \cite{Suzuki-2003}.

\section{Discrete Moyal bracket}
\label{MoyalBracket}
Using the kinetic term (the first term in the RHS of Eq. \eqref{eq:QMEWigner}) as an example, here we demonstrate the evaluation of the discrete Moyal bracket defined as Eq. \eqref{eq:Moyal}. 
The kinetic energy in a finite Hilbert space representation is expressed as 
\begin{align}
\bm{T}(p_j, q_k) &= \frac{\hbar^2}{dx^2} \left(1 - \sum_{l = -N}^{N} \exp \left(i \frac{-2 q_k (p_j - p_l)}{\hbar} \right) \langle P, l | \cos \left( \frac{\hat{p} dx}{\hbar} \right) | P, 2 j - l \rangle  \right)\notag \\
&= \frac{\hbar^2}{dx^2} \left(1 - \sum_{l = -N}^{N} \exp \left(i \frac{-2 q_k (p_j - p_l)}{\hbar} \right) \cos \left( \frac{p_l dx}{\hbar} \right) \delta_{l, 2j - l }  \right) \notag \\
&= \frac{\hbar^2}{dx^2} \left(1 - \cos \left( \frac{p_j dx}{\hbar} \right) \right).
\end{align}

Because the Moyal bracket with $\bm{A_1} = \hbar^2/dx^2$ and $\bm{A_2} = \bm{W}$ is zero, we focus on the $\cos \left( {p_j dx}/{\hbar} \right)$ term.  
Let $\bm{A_1} = \exp\left(\pm i{p_j dx}/{\hbar}\right)$ and $\bm{A_2} = \bm{W}$ in Eq. \eqref{eq:Moyal}.  Then we have
\begin{align}
&\left[\exp\left(\pm i\frac{p_j dx}{\hbar}\right) \star \bm{W} \right](p_j, q_k) = \frac{1}{(2 N + 1)^2} \sum_{j_1, j_2, k_1, k_2 = -N}^N \exp \left( i \frac{2 p_{j_2} q_{k_1} - 2 p_{j_1} q_{k_2}}{\hbar} \right) \notag \\
&\times  \exp\left(\pm i\frac{(p_j + p_{j_1}) dx}{\hbar}\right) \bm{W}( p_j + p_{j_2}, q_k + q_{k_2}) \notag \\
&= \frac{1}{(2 N + 1)} \sum_{j_1, k_2 = -N}^N \exp \left( i \frac{(\pm 1 - 2 k_2) p_{j_1} dx}{\hbar} \right) \exp\left(\pm i\frac{p_j dx}{\hbar}\right) \bm{W}( p_j , q_k + q_{k_2}) \notag \\
&= \sum_{k_2 = -N}^N \delta'_{\pm 1 - 2 k_2, 0} \exp\left(\pm i\frac{p_j dx}{\hbar}\right)\bm{W}( p_j , q_k + q_{k_2}) \notag \\
&= \exp\left(\pm i\frac{p_j dx}{\hbar} \right) \bm{W}( p_j , q_{k \pm (N + 1)})
\end{align}
Similarly, for $\bm{A_1} =  \bm{W}$ and $\bm{A_2} =  \exp\left(\pm i{p_j dx}/{\hbar}\right)$, we have
\begin{align}
&\left[ \bm{W} \star \exp\left(\pm i\frac{p_j dx}{\hbar}\right) \right] (p_j, q_k) = \exp\left(\mp i\frac{p_j dx}{\hbar}\right) \bm{W}( p_j , q_{k \pm (N + 1)}).
\end{align}
Thus the discrete Moyal product of the kinetic energy is expressed as
\begin{align}
&-\frac{i}{\hbar}[ \bm{T} \star \bm{W}](p_j, q_k) = - \hbar \sin \left(\frac{p_j dx}{\hbar} \right) \frac{\bm{W}( p_j , q_{k + N + 1}) - \bm{W}( p_j , q_{k - N - 1})}{dx^2}.
\label{eq:advection}
\end{align}
% For example, for $k=0$, i. e. $q_0=0$, the finite difference of the above expression involves the elements.  
For example, for $q_0$, the above expression involves the contributions from $q_{N + 1 \equiv -N (mod \ 2N + 1)}$ and $q_{-N - 1 \equiv N (mod \ 2N + 1)}$, which are the elements near the boundary of the periodic state.
Note that $N + 1$ arises from $\delta'_{1 - 2 k_2, 0}$ that is the inverse element of 2 modulo $2N + 1$. 
For large $N$, the above expression reduces to the kinetic term of the conventional QFPE by regarding the finite difference near the  boundary as the derivative of the coordinate.

\section{Discrete quantum Fokker-Planck equation for large $N$}
\label{QFPELargeN}

For a large $N$, Eq. \eqref{eq:QMEWignerDIS} reduces to
\begin{align}
\frac{\partial}{\partial t} W(p, q) &= -p \frac{\partial}{\partial q} W(p, q) -\frac{i}{\hbar} \{ \bm{U} , \bm{W} \}_M + \frac{\eta}{ \beta} \frac{\partial^2}{\partial p^2}W(p, q) \notag \\
&+\frac{\eta}{2} \left(\hat{M}^2_p \hat{M}_x W(p, q) + p \hat{M}_x \frac{\partial}{\partial p} W(p, q)  \right),
\label{eq:QMEKramers}
\end{align}
where 
\begin{align}
\frac{\partial W(p, q)}{\partial q} \equiv \frac{W(p_j, q_{k + N + 1)} - W(p_j, q_{k - N - 1})}{dx},
\end{align}
\begin{align}
\frac{\partial  W(p, q)}{\partial p} \equiv \frac{W(p_{j + N + 1}, q_k) - W(p_{j - N - 1}, q_k)}{dp},
\end{align}
\begin{align}
\hat{M}_x W(p, q)  \equiv \frac{W(p_j, q_{k + N + 1)} + W(p_j, q_{k - N - 1})}{2} ,
\end{align}
and
\begin{align}
\hat{M}_p W(p, q)  \equiv \frac{W(p_{j + N + 1}, q_k) + W(p_{j - N - 1}, q_k)}{2}.
\end{align}
Although the above expression has a similar form to the QFPE, the finite difference operators for the discrete WDF are defined by the elements near the periodic boundary, i.e., for $W(p_0, q_0)$, $\partial /\partial q$ is evaluated 
from  $W(p_0, q_{-(N+1)} )$, and $W(p_0, q_{N+1})$. Thus the appearance of the discrete WDF can be different from the regular WDF, as depicted in Fig. \ref{eqHoWDF}  even for large $N$.



\end{document}